\documentclass[12pt]{iopart}
\usepackage{graphicx}

\begin{document}


\title[Modified hadronization in Au+Au collisions]{Observation of
  modified hadronization in relativistic Au+Au collisions: a promising
  signature for deconfined quark-gluon matter }

\author{Paul Sorensen}

\address{Lawrence Berkeley National Laboratory, One Cyclotron Road, Berkeley, California 94720}
\ead{prsorensen@lbl.gov}

\begin{abstract} 
Measurements of identified particles from Au+Au collisions at
$\sqrt{s_{_{NN}}}=200$~GeV are reviewed. Emphasis is placed on nuclear
modification, baryon-to-meson ratios, and elliptic flow at
intermediate transverse momentum ($1.5 < p_T < 5$ GeV/c). Possible
connections between (1) these measurements, (2) the running coupling
for static quark anti-quark pairs at finite temperature, and (3) the
creation of a deconfined quark-gluon phase are presented.
Modifications to hadronization in Au+Au collisions are proposed as a
likely signature for the creation of deconfined colored matter.
\end{abstract}



\submitto{\JPG} 

\section{Introduction}\label{intro}

A central question to be answered by scientists at Brookhaven National
Laboratory's Relativistic Heavy Ion Collider (RHIC) is: if nuclei are
collided together with high enough center-of-mass energy, is it possible
to create matter in which quarks and gluons become deconfined ---
\textit{i.e.} are colored objects able to roam over distances greater
than hadronic length scales~\cite{RHIC}?

\begin{figure}[htb]
\vspace*{-.4cm}
\resizebox{.5\textwidth}{!}{\includegraphics{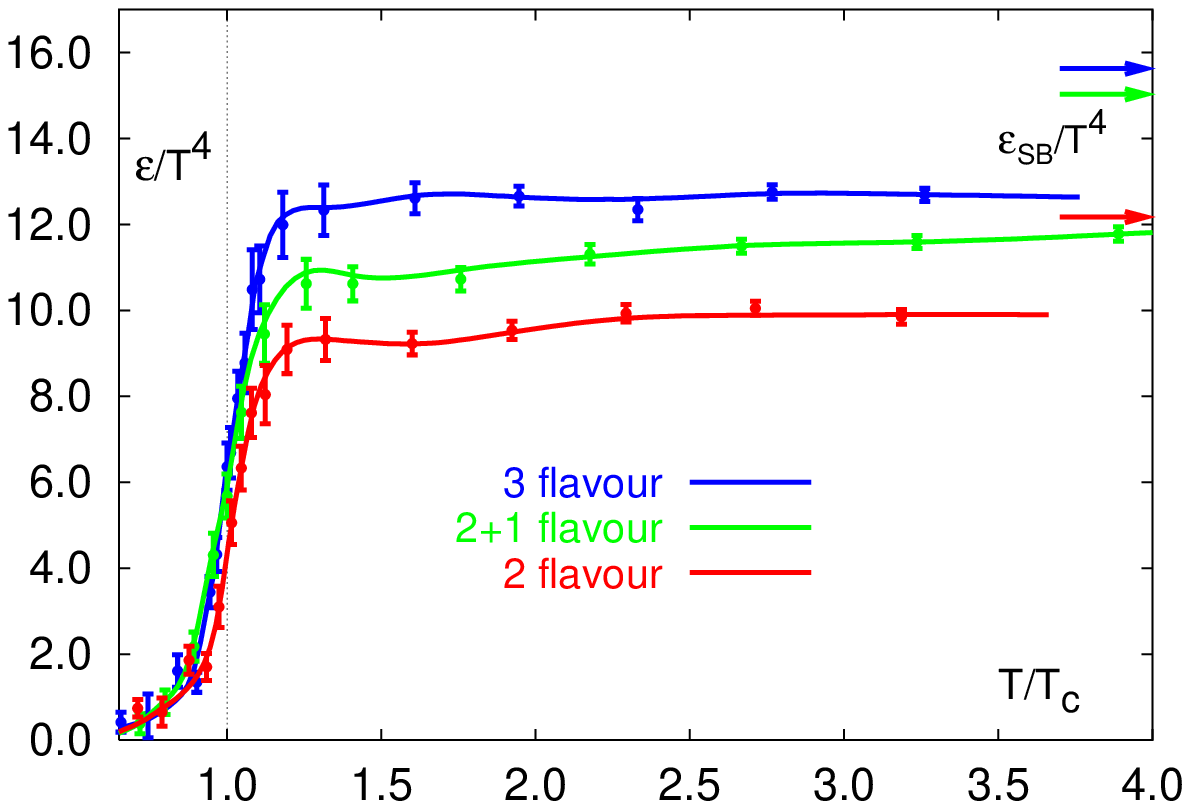}}
\resizebox{.5\textwidth}{!}{\includegraphics{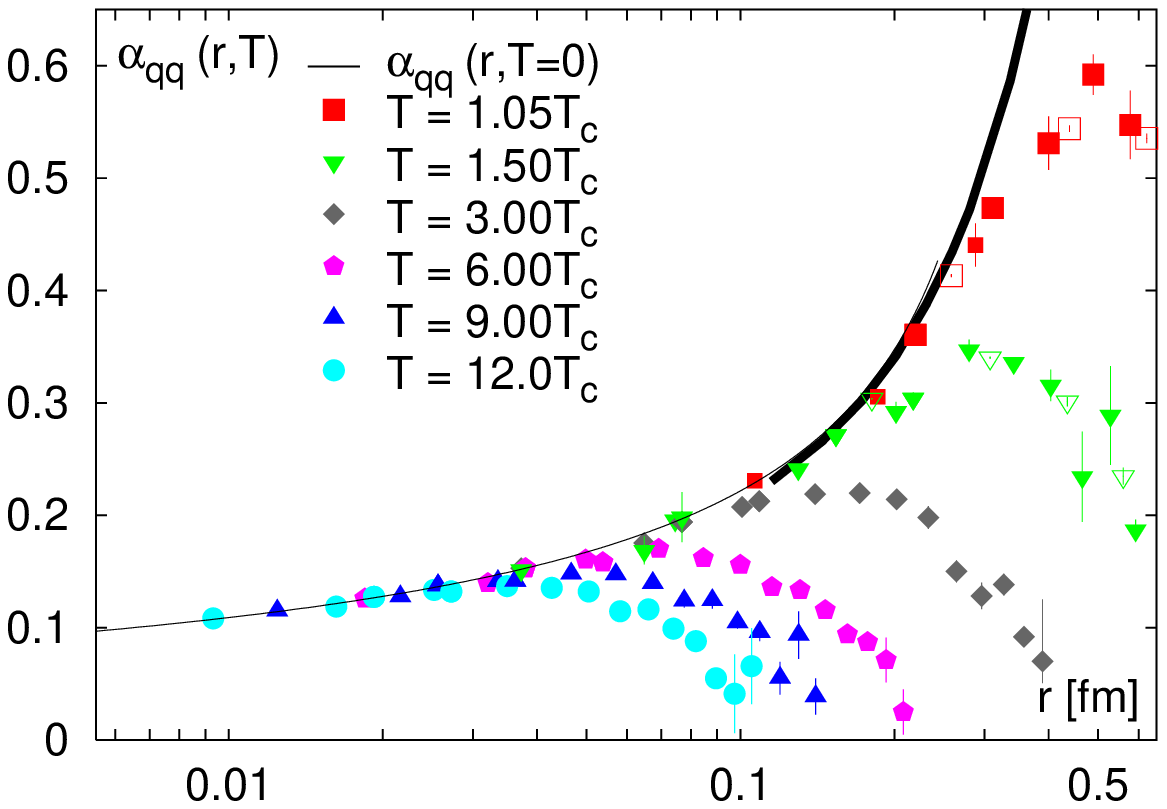}}
\vspace*{-.7cm}
\caption[]{ Left: Lattice QCD calculations of the dependence of the
  energy density $\epsilon$ scaled by $T^4$ on $T/T_c$. Right: Lattice
  QCD calculations of the running coupling of static, heavy quark
  anti-quark pairs at finite temperature as a function of their
  separation $r$.}
\label{lattice}
\end{figure}

Lattice QCD calculations demonstrate that as the temperature of
hadronic matter is increased above a critical value ($T_c$) quark and
gluon degrees-of-freedom become accessible. This is most graphically
demonstrated by the results from Ref.~\cite{Karsch:2001cy} shown in
Fig.~\ref{lattice} (left) where the energy density $\epsilon$ scaled
by $T^4$ is shown for different values of $T/T_c$. Calculations of the
quark anti-quark coupling $\alpha_{qq}$ indicate, however, that even
at temperatures several times greater than $T_c$, quarks and
anti-quarks still feel effects of confinement~\cite{Kaczmarek:2004gv}:
\textit{i.e.} ``remnants of confinement''.  Fig.~\ref{lattice} (right)
shows the QCD coupling constant $\alpha(r,T/T_c)$ for static, heavy
quark anti-quark pairs as a function of their separation $r$ for
temperatures from $1.05\cdot T_c$ to $12\cdot T_c$. High-statistics
lattice results from Ref.~\cite{Necco:2001xg} for the coupling at
$T=0$ are shown on the figure as a thick, black line.


Measurements indicate that in Au+Au collisions at top RHIC energy, the
energy density reaches values well above the critical energy density
for deconfinement~\cite{mult}. The fireball formed in these collisions
rapidly expands and cools. If matter is formed with $T>T_c$, these
calculations suggest that as the matter cools and approaches $T_c$
from above, the forces of confinement will turn on gradually. In this
case, the process by which quarks and gluons become confined
(\textit{i.e.}  hadronization) may be significantly modified in
comparison to cases where a deconfined matter is not formed --- in
which case hadronization will take place in vacuum and
$\alpha_{qq}(r)$ will have the form corresponding to the $T=0$ case
shown as a thick, black line in Fig.~\ref{lattice}. Although the
finite temperature calculations have only been made for static, heavy
quark anti-quark pairs, the modifications to $\alpha(r,T/T_c)$ are
likely to be robust and also present in the case of light quarks. As
such, \textbf{the best way to detect the presence of deconfined,
  colored matter in ultra-relativistic nuclear collisions \textit{may
    be} to study how hadron formation is modified in Au+Au collisions
  compared to more elementary collisions} (\textit{e.g.} $e^++e^-$ or
$p+p$ collisions).

According to de Broglie's relation
($p=h/\lambda$)
the momentum $p$ of a probe and the corresponding length scale that it
can resolve (\textit{i.e.} the wavelength $\lambda$) are inversely
related. As the momentum transfer involved in scatterings increases,
the resolution with which matter can be probed also increases. In
nucleus-nucleus collisions at RHIC it is not feasible to probe the
collision system with an external probe (as in scattering electrons or
protons off the system formed after the gold ions collide). Instead,
we rely on the particles produced by the collision to probe the matter
formed.

For particles produced with sufficiently large momentum, a picture of
the collision system as being composed of point-like particles
(partons) should hold. Most particles produced in nucleus-nucleus
collisions, however, are at much lower momentum and their long
\textit{wavelengths} probe the bulk characteristics of the produced
matter (\textit{i.e.} temperature, collective velocity, system-sizes,
etc.~\cite{Adler:2001nb,Adler:2001zd}). For momenta between the two
extremes (\textit{e.g.} above 1 GeV/c), one might expect to probe
length scales that are commensurate with the systems constituents ---
whether hadrons, ``remnants of confinement'', or constituent quarks
--- without yet reaching the limit where the most appropriate
description of the system is that of point-like partons. If the
momentum is too high, we'll probe the region of $\alpha(r,T/T_c)$
that, according to Fig.~\ref{lattice}, is independent of
temperature. If the momentum is too low, we may not have sensitivity
to the modifications shown in the figure. For these reasons, the
kinematic range where we would most naturally expect to find evidence
for modifications to hadron formation is just below the momentum scale
where a partonic, point-like picture holds.

\section{Observations at intermediate-$p_T$}

Three measurements from RHIC in the intermediate $p_T$ range
(1.5~GeV/c $< p_T < 5$~GeV/c) point towards modifications to the
process of hadron formation in Au+Au collisions:
\begin{itemize}
  \item the nuclear modification factors --- \textit{i.e.} the ratio
    of yields in central Au+Au collisions compared to peripheral Au+Au
    ($R_{CP}$) or p+p ($R_{AA}$) collision scaled by the number of
    binary nucleon-nucleon collisions~\cite{Adler:2003kg,Adams:2003am}.
  \item the baryon-to-meson ratios~\cite{Adler:2003kg,hlmsb}.
  \item the anisotropy in azimuth of particle production relative to
    the reaction plane --- \textit{i.e.} the elliptic flow parameter
    $v_2$~\cite{Adams:2003am,Adler:2003kt,sorensen,msbv2,Adams:2004bi}.
\end{itemize}

\begin{figure}[hbtp]
\centering\mbox{
\includegraphics[width=0.65\textwidth]{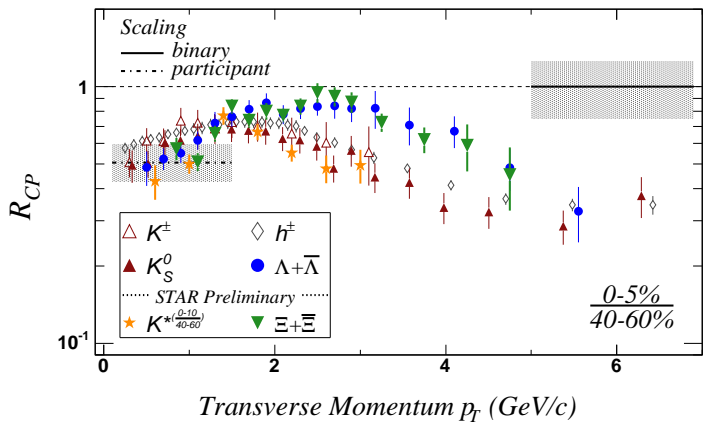}
}
\caption{ Nuclear modification factors ($R_{CP}$) for various
  identified particles measured in Au+Au collisions at
  $\sqrt{s_{_{NN}}}=200$~GeV by the STAR collaboration. } \label{rcp}
\end{figure}

Fig.~\ref{rcp} shows the nuclear modification factor $R_{CP}$ measured
with charged hadrons~\cite{highpt}, charged kaons, $K_S^0$s,
$K^*$(892)s~\cite{Zhang:2004rj}, $\Lambda$s$+\overline{\Lambda}$s, and
$\Xi$s$+\overline{\Xi}$s~\cite{hlmsb}. In the case that the centrality
dependence of particle yields scales with the number of binary
nucleon-nucleon collisions, $R_{CP}$ will equal one. At $p_T=5$~GeV/c,
charged hadron yields in central collisions are suppressed from
expectations by a factor of four. This suppression is taken as a
signature of the quenching of jets in the bulk matter formed in
central collisions. The charged and neutral kaons show a suppression
and $p_T$ dependence similar to that of charged hadrons. By contrast,
the $R_{CP}$ values for $\Lambda$s$+\overline{\Lambda}$s and
$\Xi$s$+\overline{\Xi}$s rise above the charged hadron values at
$p_T\approx 1.5$~GeV/c and approach unity. They remain above charged
hadron and kaon $R_{CP}$ values until $p_T\approx 5$--6~GeV/c. Similar
observations were made by the PHENIX collaboration with proton
$R_{AA}$ rising above pion $R_{AA}$ in a similar $p_T$
region~\cite{Adler:2003kg}.

\begin{figure}[htb]
\vspace*{-.0cm}
\resizebox{.595\textwidth}{!}{\includegraphics{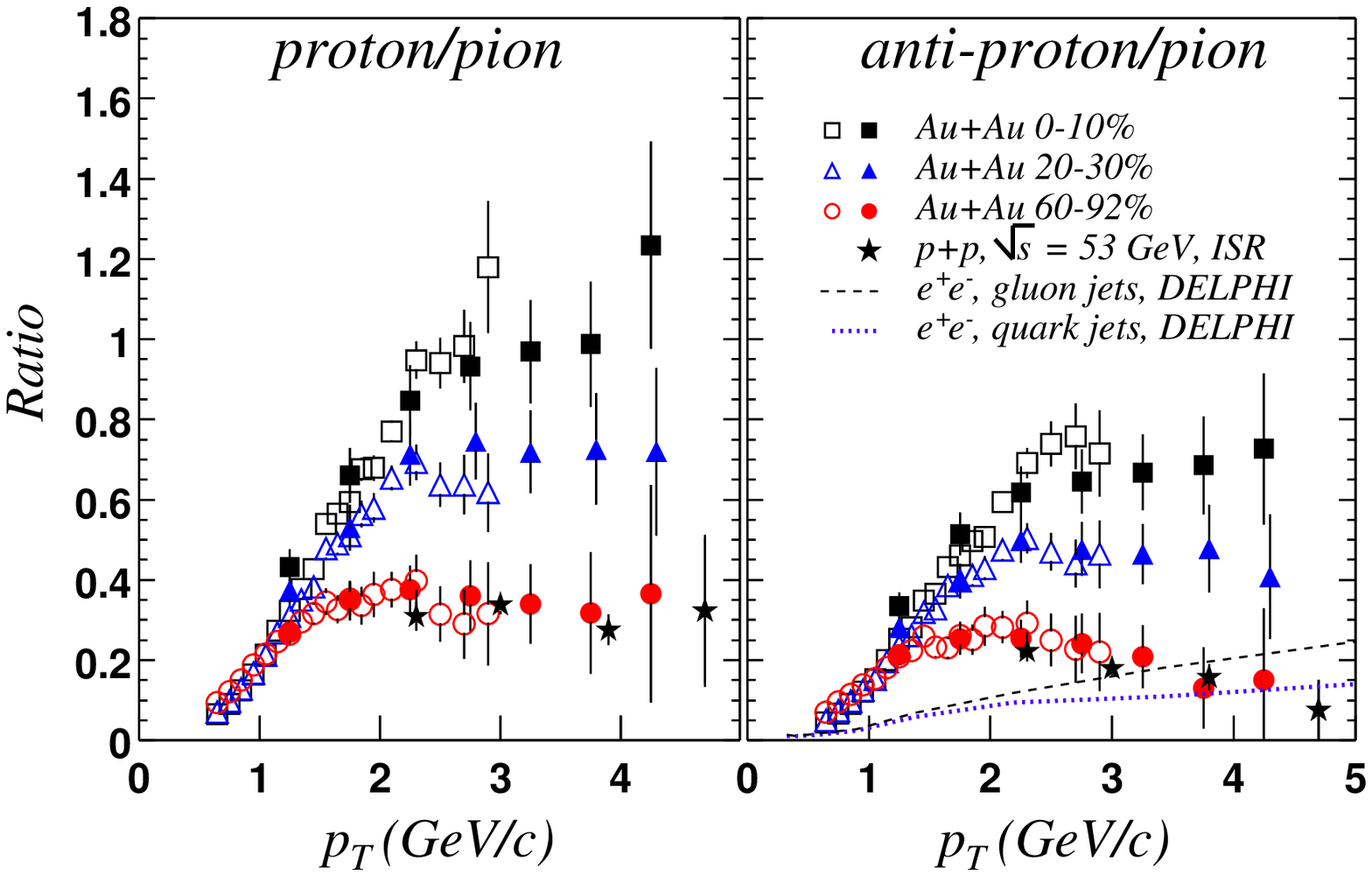}}
\resizebox{.405\textwidth}{!}{\includegraphics{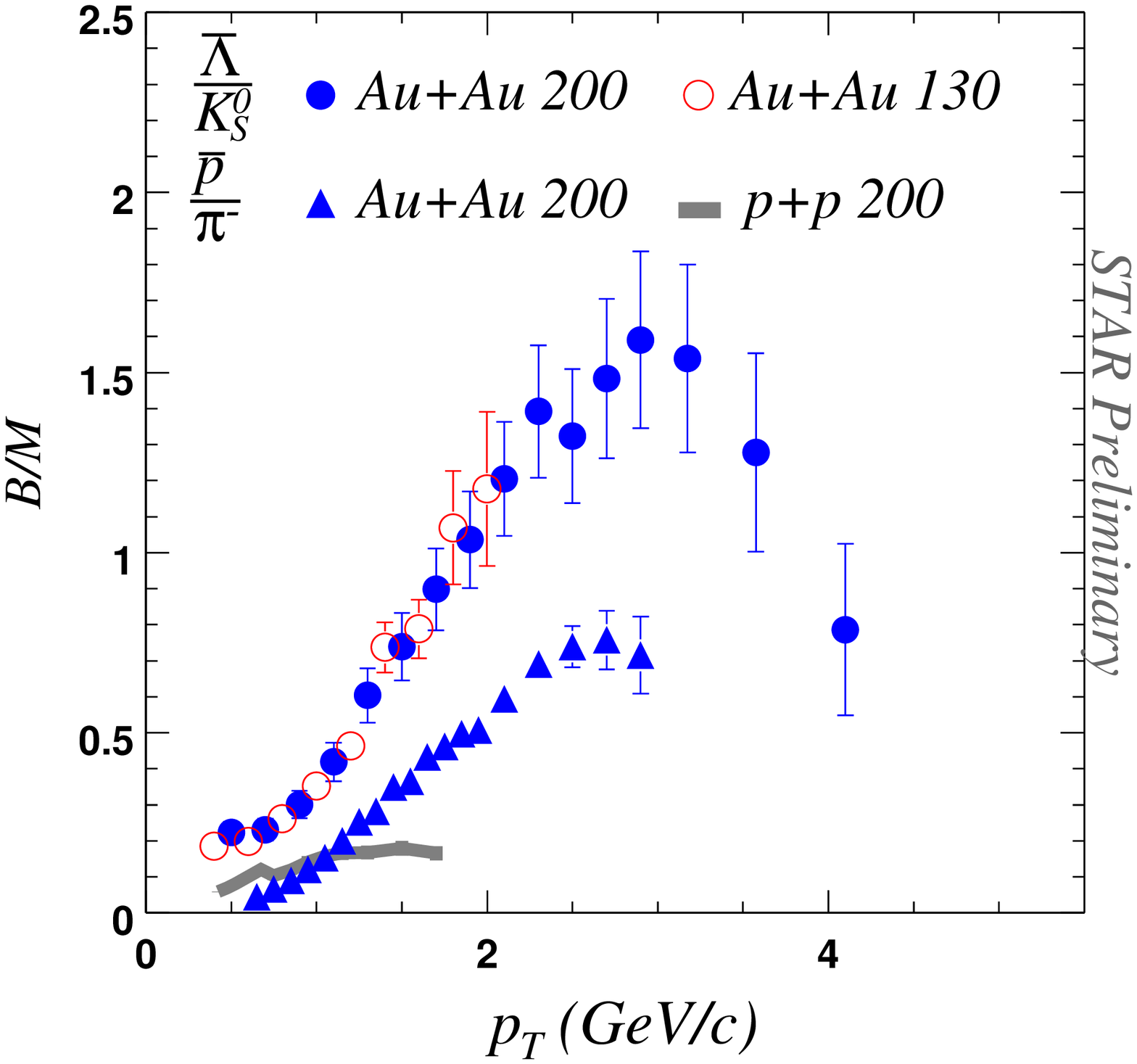}}
\vspace*{-.4cm}
\caption{ Left: The proton/pion ratios measured in Au+Au collisions
  at $\sqrt{s_{_{NN}}}=200$~GeV by PHENIX, compared to measurements
  from $e^++e^-$ and $p+p$ collisions. Right:
  $\overline{\Lambda}$/$K^0_S$ ratios for central Au+Au collisions at
  $\sqrt{s_{_{NN}}}=200$~GeV and 130~GeV measured by STAR. The
  $\overline{p}/\pi^-$ ratio from p+p collisions from STAR are also
  shown.} \label{B/M}
\end{figure}

$R_{CP}$ values for baryons being greater than those of mesons
signifies that baryon production increases with centrality (and
therefore collision overlap-density) more quickly than meson
production. When $R_{CP}$ or $R_{AA}$ values for different particle
types take on the same value (\textit{e.g.} at $p_T=5$--6~GeV/c), it
means that the relative abundances of the different particles matches
that observed in $p+p$ collisions. As such, the baryon-to-meson ratios
in central Au+Au collisions at intermediate $p_T$, must also reflect
the same baryon excesses. Fig.~\ref{B/M} shows measurements of
proton/pion, anti-proton/pion~\cite{Adler:2003kg} and
$\overline{\Lambda}$/$K_S^0$~\cite{hlmsb} ratios as a function of
$p_T$ for various centralities and collision systems. At intermediate
$p_T$, a striking difference is observed between the baryon-to-meson
ratios in central Au+Au collisions and those in
$e^++e^-$~\cite{Abreu:2000nw} or $p+p$
collisions~\cite{Alper:1975jm}. The measurements in Figs.~\ref{rcp}
and \ref{B/M} demonstrate that the process by which parton
distributions are mapped onto hadron distributions is drastically
different in Au+Au and $p+p$ collisions. As long as hadronization is
not modified, a change to the underlying parton distributions will not
lead to such a drastic change in the relative abundances.

\begin{figure}[hbtp]
\centering\mbox{
\includegraphics[width=0.65\textwidth]{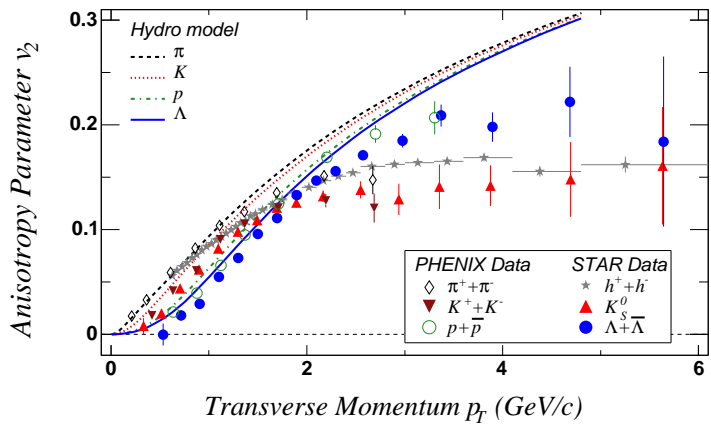}
}
\caption{ $v_2$ for a variety of particles from a minimum-bias sample
  of Au+Au collisions at $\sqrt{s_{_{NN}}}=200$~GeV measured by the
  STAR~\cite{Adams:2003am} and PHENIX~\cite{Adler:2003kt}
  collaborations. Curves show the results from hydrodynamic model
  calculations~\cite{hydro}. }
\label{v2}
\end{figure}

In non-central nucleus-nucleus collisions, the collision overlap
region is elliptic in shape. Secondary interactions, can convert this
initial coordinate-space anisotropy to an azimuthal anisotropy in the
final momentum-space distributions. That anisotropy is commonly
expressed in terms of the coefficients from a Fourier expansion of the
azimuthal component of the invariant yield~\cite{flow}. The second
component (the ``elliptic flow'' parameter $v_2$) is the largest and
most commonly studied of the coefficients. Fig.~\ref{v2} shows $v_2$
for pions, kaons, protons and Lambda
hyperons~\cite{Adams:2003am,Adler:2003kt}.  In the low momentum region
($p_T<1.5$~GeV/c), $v_2$ is increasing with
$p_T$~\cite{Ackermann:2000tr}. In this region, for a given $p_T$, the
$v_2$ values are ordered by mass; with more massive particles having
smaller $v_2$ values~\cite{Adams:2003am,Adler:2003kt,sorensen,pidv2}.
This mass ordering is reasonably well described by the hydrodynamic
model calculations~\cite{hydro} also shown in the figure.

For $p_T > 1.5$~GeV/c, The data and model calculations deviate. The
model over-predicts the measured $v_2$ values and the particle-type
dependence reverses: the $v_2$ values for the more massive baryons are
larger than those for the mesons. A model that relies on partonic,
in-medium, energy-loss and vacuum fragmentation will predict smaller
$v_2$ values when $R_{CP}$ is closer to one~\cite{dedx}. The data,
however, indicate that although protons, $\Lambda$s, and
$\overline{\Lambda}$s have $R_{CP}$ values near unity, their maximum
$v_2$ values exceed those of pions and kaons by approximately
50\%. Models that assume hadrons form via the coalescence of co-moving
constituent-quarks, however, are able to simultaneously reproduce the
particle-type dependencies of $R_{CP}$ and $v_{2}$~\cite{reco}. By
extension, these models also account for the anomalously large
baryon-to-meson ratios for central Au+Au collisions.

\begin{figure}[hbtp]
\centering\mbox{
\includegraphics[width=0.65\textwidth]{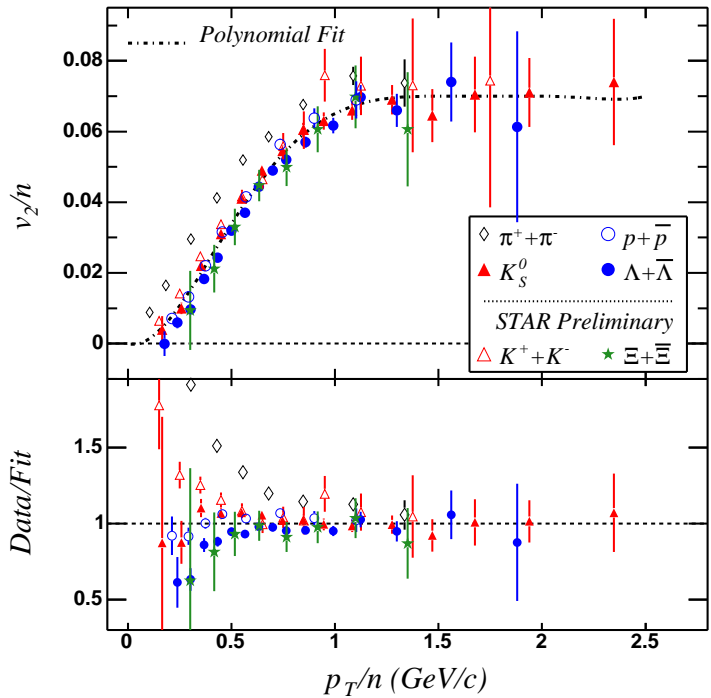}
}
\caption{ Quark number scaled $v_2$ ($v_{2}(p_T/n_q)/n_q$) for Au+Au
  collisions at $\sqrt{s_{_{NN}}}=200$~GeV from
  Ref.~\cite{Adams:2004bi}. The bottom panel shows the ratio of the
  data to the fit. Pions are excluded from the fit but are shown in
  the figure for comparison. } \label{v2scaling}
\end{figure}

Coalescence models for hadronization predict that at intermediate
$p_T$, $v_2$ for different hadrons will follow a scaling law based on
the number of valence quarks in the corresponding hadron ($n_q$):
\begin{equation}
v_2^{h}(n_q\cdot p_T^{q})=n_q\cdot v_2^{q}(p_T^{q}),
\end{equation}
where $v_2^h$, $v_2^q$, $p_T^h$, and $p_T^q$ are hadron $v_2$, quark
$v_2$, hadron $p_T$, and quark $p_T$ respectively. If this scaling
holds then $v_2^h(p_T^h/n_q)/n_q$ should fall on a universal curve. In
these models this universal curve represents the $v_2$ value for the
quark distributions prior to hadronization. In Fig.~\ref{v2scaling}
this scaling is tested by plotting $v_2$ versus $p_T$, where both axes
have been scaled by $n_q$. A polynomial curve is fit to all the data
and the ratio of the data to the universal curve is plotted in the
bottom panel. Very good agreement with the scaling prediction is found
for all the hadron species except pions. The pion $v_2$ values deviate
from the universal curve throughout the measured $p_T$ range and are
not included in the fit. This deviation does not necessarily signal a
violation of the scaling law because, the pion sample used to
calculate $v_2$ is dominated by pions from resonance decays. Although
the effect of resonance decays on the $v_2$ for other hadrons has been
found to be relatively small, the effect was found to significantly
increase the measured pion $v_2$~\cite{decayv2}.

The larger rate-of-increase for baryon production (\textit{i.e.}
$R_{CP}^{\mathrm{Baryon}} > R_{CP}^{\mathrm{Meson}}$), the anomalously
large baryon-to-meson ratios in Au+Au collisions, and the
quark-number-scaling for $v_2$, point to possible modifications of
hadron formation in Au+Au collisions compared to p+p
collisions. Models in which hadronization occurs via coalescence of
co-moving constituent quarks successfully account for these three
observations~\cite{reco}. These hadronization models demonstrate one
way that the hadron formation process could be modified. Whether these
models remain successful as more precise measurements are made is
still an open question. The evidence for modifications to
hadronization from data, however, are clear and independent of the
theoretical models.

\section{Conclusions}\label{concl}

Identified particle nuclear modification $R_{CP}$ and elliptic flow
$v_2$ measurements made in Au+Au collisions at RHIC have revealed an
apparent quark-number dependence in the $p_T$ region from 1.5~GeV/c to
5~GeV/c. The $v_2$ measurements follow a quark-number scaling
predicted by models of hadron formation via coalescence of co-moving
quarks. The $R_{CP}$ measurements show that baryon production
increases more rapidly with centrality than meson production --- an
observation that also supports a picture of hadron formation by quark
coalescence or recombination. These observations are consistent with
the presence of modifications to the hadronization process in Au+Au
collisions compared to $e^++e^-$ or $p+p$ collisions.

A picture of hadron formation with an intermediate, constituent-quark
step, is highly suggestive of the formation of deconfined quark-gluon
matter. The modified hadron formation process points to a scenario
where matter is created with a temperature above $T_c$ and then cools
--- approaching $T_{c}$ from above --- until confinement is gradually
re-enforced. Lattice calculations of the QCD coupling constant
$\alpha(r,T/T_{c})$ for static quark anti-quark pairs show that this
process is consistent with the equations of QCD. 
\textbf{These calculations and the empirical observations presented
  here demonstrate that modifications to the hadronization process in
  nuclear collisions are promising signatures for the creation of
  deconfined quark-gluon matter.}

\section*{Acknowledgements}
The author thanks the conference organizers and acknowledges valuable
input from H.~Huang, K.~Schweda, R.~Seto, and N.~Xu.

\vfill\eject
\end{document}